\documentclass[a4paper,11pt]{article}
\usepackage[english]{babel}
\usepackage[pdftex]{graphicx}

\begin{document}
\pagenumbering{arabic}

\newcommand{\KPPEE}{$K^{\pm} \to \pi^{\pm}\pi^{0}e^+e^-$}
\newcommand{\KPPGS}{$K^{\pm} \to \pi^{\pm}\pi^{0}\gamma^*$}
\newcommand{\PPEE}{$\pi^{\pm}\pi^{0}e^+e^-$}

\title{NA48/2 studies of rare decays}

\author{M.~Raggi\footnote{for the NA48/2 Collaboration}\hspace{1mm}\footnote{corresponding author: mauro.raggi@lnf.infn.it}\\
{\it Laboratori Nazionali di Frascati - INFN, Frascati (Rome), Italy } \\
}
\date{to be published in the La Thuile 2015 proceedings}

\maketitle

\begin{abstract}
The first observation of about 2000 candidates, with a background contamination
below 3\%, of the rare decay \KPPEE is reported by the NA48/2 experiment. 
The preliminary branching ratio in the full kinematic region is obtained to be: ${\cal B}(K^{\pm} \to \pi^{\pm}\pi^{0}e^+e^-)=(4.06\pm0.17)\cdot10^{-6}$ 
by analyzing the data collected in 2003.
A sample of $4.687\times 10^6$ $K^\pm\to\pi^\pm\pi^0_D$, decay candidates with a negligible background contamination collected in 2003--04 is analyzed to search for the dark photon ($A'$) via the decay chain $K^\pm\to\pi^\pm\pi^0$, $\pi^0\to\gamma A'$, $A'\to e^+e^-$. No signal is observed, and preliminary limits in the plane dark photon
mixing parameter $\varepsilon^2$ versus its mass $m_{A'}$ are reported.

\end{abstract}

\section{The NA48/2 experiment}
\label{sec:experiment}
The NA48/2 experiment at the CERN SPS collected a large sample of charged kaon ($K^\pm$) decays during its 2003-04 data taking period.
The NA48/2 beam line has been designed to deliver simultaneous narrow momentum band $K^+$ and $K^-$ beams originating from the collision of the primary 400
GeV/$c$ protons extracted from the CERN SPS on a beryllium target. Secondary beams with central momenta of $(60\pm3)~{\rm GeV}/c$ (r.m.s.) following a common beam axis were used. The beam kaons decayed in a fiducial
decay volume contained in a 114~m long cylindrical vacuum tank. The momenta of charged decay products were measured in a magnetic spectrometer, housed in a tank filled
with helium placed after the decay volume. The spectrometer comprised four drift chambers (DCHs) and a dipole magnet. A plastic scintillator
hodoscope (CHOD) producing fast trigger signals and providing precise time measurements of charged particles was placed after the spectrometer. Further downstream was a
liquid krypton electromagnetic calorimeter (LKr), an almost homogeneous ionization chamber with an active volume of 7~m$^3$ of liquid krypton, $27X_0$ deep, segmented
transversally into 13248 projective $\sim\!2\!\times\!2$~cm$^2$ cells and with no longitudinal segmentation. An iron/scintillator hadronic calorimeter and muon detectors were located further downstream. A dedicated two-level trigger was used to collect three
track decays with a very high efficiency. A detailed description of the detector can be found in \cite{fa07}.

\section{First observation of \KPPEE decay}
The \KPPEE decay proceeds through virtual photon exchange which undergoes internal
conversion into electron-positron pair, i.e. \KPPGS $\to$ \PPEE.  The $\gamma^*$ is produced by two different mechanisms: Inner Bremsstrahlung (IB), where the $\gamma^*$ is
emitted by one of the charged mesons in the initial or final state and Direct Emission (DE) when $\gamma^*$ is radiated off at the weak
vertex of the intermediate state. As a consequence the  differential decay width consists of three
terms: the dominant long-distance IB contribution (pure electric part E), the DE component
(electric E and magnetic M parts) and the interference between them\cite{Cappiello:2011qc}. The interference term collects the different contributions, IBE, IBM and EM. 
For this reason the \PPEE decay offers interesting short and long distance parity violating observables. In the $K^{\pm} \to 
\pi^{\pm}\pi^{0}\gamma$ mode the interference consists only of the IBE term\cite{Christ:1967zz}, because the remaining (EM) interferences are P-violating, but cancel out upon angular integration. There are few theoretical publications related to the \KPPEE \cite{Cappiello:2011qc}\cite{Pichl:2000ab}\cite{Gevorkyan:2014waa}.
Recently authors of $\cite{Cappiello:2011qc}$ where able to predict, on the basis of the NA48/2 measurement of the magnetic and electric terms in $K^{\pm} \to \pi^{\pm}\pi^0\gamma$ \cite{Batley:2010aa}, the branching ratio of the single components. No experimental observation has so far been reported. 
\subsection{Selection and background estimates}
\KPPEE event candidates are reconstructed from three charged tracks and two photons, forming neutral pion,
pointing to a common vertex in the fiducial decay volume. Particle identification is based on the energy deposition  in LKr (E) associated or not to a charged track momentum (p) measured in the spectrometer. The charged track is identified as electron/positron if its E/p ratio is
greater than 0.85, and as a charged pion if the E/p ratio is lower than 0.85. Two isolated energy
clusters without associated track in the LKr are identified as the two candidates photons
from the $\pi^0$ decay. Their invariant mass is required to be within $\pm$10 MeV/$c^2$ from the nominal PDG\cite{pdg} $\pi^0$ mass.
The reconstructed invariant mass of the \PPEE system is required to be within $\pm$10 MeV/$c^2$ from the nominal PDG\cite{pdg} $K^{\pm}$ mass.
Two main sources of background are contribution to the signal final state: $K^{\pm} \to \pi^{\pm}\pi^{0}\pi^{0}_{D}$ ($K_{3\pi D}$) when one of the photon is lost, and $K^{\pm} \to \pi^{\pm}\pi^{0}_{D}
(\gamma)$ ($K_{2\pi D}$), where $\pi^{0}_{D}$ denotes the $\pi^{0}$ Dalitz decay $\pi^{0} \rightarrow e^+ e^- \gamma$. The suppression of the $K_{3\pi D}$ background events is obtained by requiring the squared invariant mass of the $\pi^+\pi^0$ system to be greater 
than 120 MeV$^2/c^4$, exploiting the presence of three particles with almost the same mass in the final state. In order to reject $K_{2\pi D}$ background contamination both the 
invariant masses $M_{ee\gamma_{1,2}}$ are required to be more than 7 MeV/$c^2$ away from the nominal mass of the neutral pion. 
Analyzing the 2003 data, a sample of 1916 signal candidates has been selected with a background contamination below 3\%.
In particular MC simulation predicts a contribution of (26$\pm$5.1) candidates form $K_{2\pi D}$  and (30$\pm$5.5)from $K_{3\pi D}$ events. 
The normalization mode ($K_{2\pi D}$) is recorded concurrently with the signal mode , using the same trigger logic. 
A common event reconstruction is considered as much as possible aiming to cancel of systematic effects such as particle 
identification and trigger inefficiencies.
The selection of the normalization mode $K_{2\pi D}$ uses the same set of requirements as the signal selection except for the $\pi^0$-reconstruction and background suppression parts. The neutral pion is reconstructed by requiring only one $\gamma$-candidate cluster and computing its invariant
mass with the electron and positron pair. The only background source for the normalization channel is the $K_{\mu3D}$ mode ($K^{\pm} \rightarrow \mu^+ \nu \pi^{0}_{D}$).
In the whole 2003 data sample 6.715 million $K_{2\pi D}$ candidates are selected with a background
contamination smaller than 0.1\%.

\subsection{Branching ratio measurement}
\begin{figure}
 \centering
 \includegraphics[width=75mm,scale=0.5]{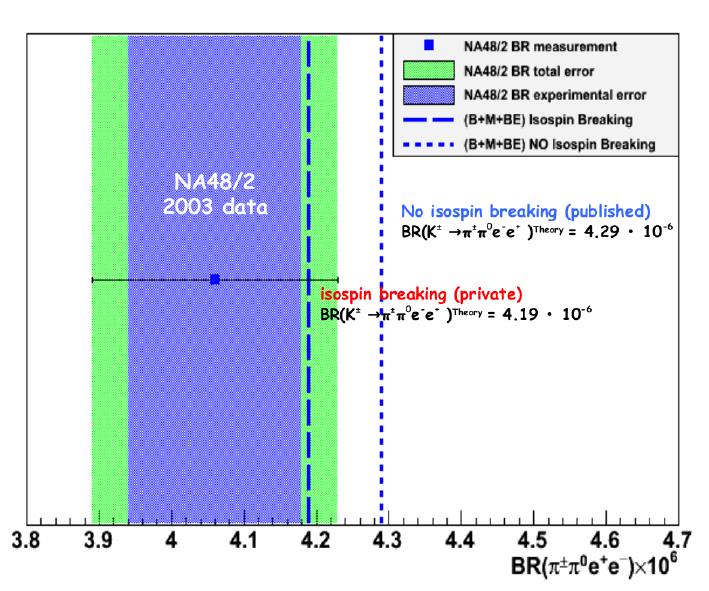}
        \caption{The \KPPEE preliminary branching ratio is plotted with
its experimental error (shaded blue band) and its total error (shaded green band).}
        \label{fig:BRppee}
\end{figure}

The total Branching Ratio of \KPPEE is obtained using the expression:
\begin{equation}
{\cal B}(K^{\pm} \to \pi^{\pm}\pi^{0}e^+e^-)=\frac{N_S-N_B}{N_N}\frac{A_N\epsilon_N}{A_S\epsilon_S}{\cal B}(N)
\end{equation} 
where $N_{S,B,N}$ are the number of signal (1916), background (55.8$\pm$7.4) and $K_{2\pi D}$ events.
$A_{S,N}$ $\epsilon_{S,N}$ are the acceptances and trigger efficiencies of the signal and normalization modes.
The normalization mode branching ratio ${\cal B}(N) = (2.425 \pm0.076)\cdot10^{-3}$ is obtained from the PDG\cite{pdg} world average.
The trigger efficiencies ($\epsilon$), very similar ($\sim$98\%) for signal and normalization mode, are measured on data using control samples.
The acceptances of the signal, the normalization and the background channels are computed
using GEANT3-based\cite{GEANT} MC simulations which include the full detector and material description,
stray magnetic fields, beam line geometry and local detector imperfections.
\begin{figure}[ht]
    \centering
    \begin{minipage}[t]{0.45\textwidth}
        \includegraphics[width=75mm,scale=0.5]{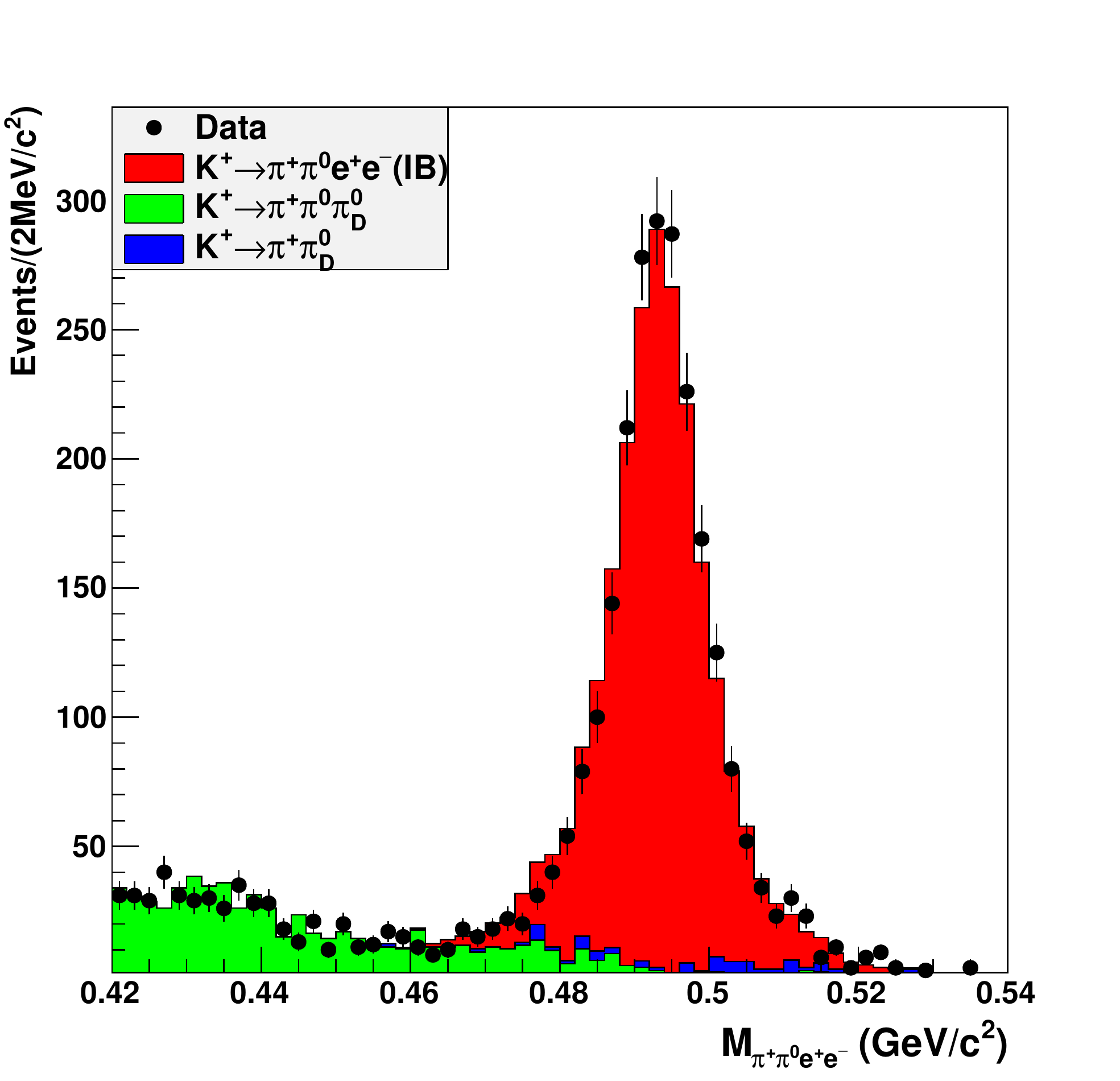}%
        \caption{Reconstructed \PPEE invariant mass distributions of the data and simulated background samples.}
        \label{fig:ppeemass}
    \end{minipage}%
    \hspace{1.cm}
    \begin{minipage}[t]{0.45\textwidth}
        \includegraphics[width=75mm,scale=0.5]{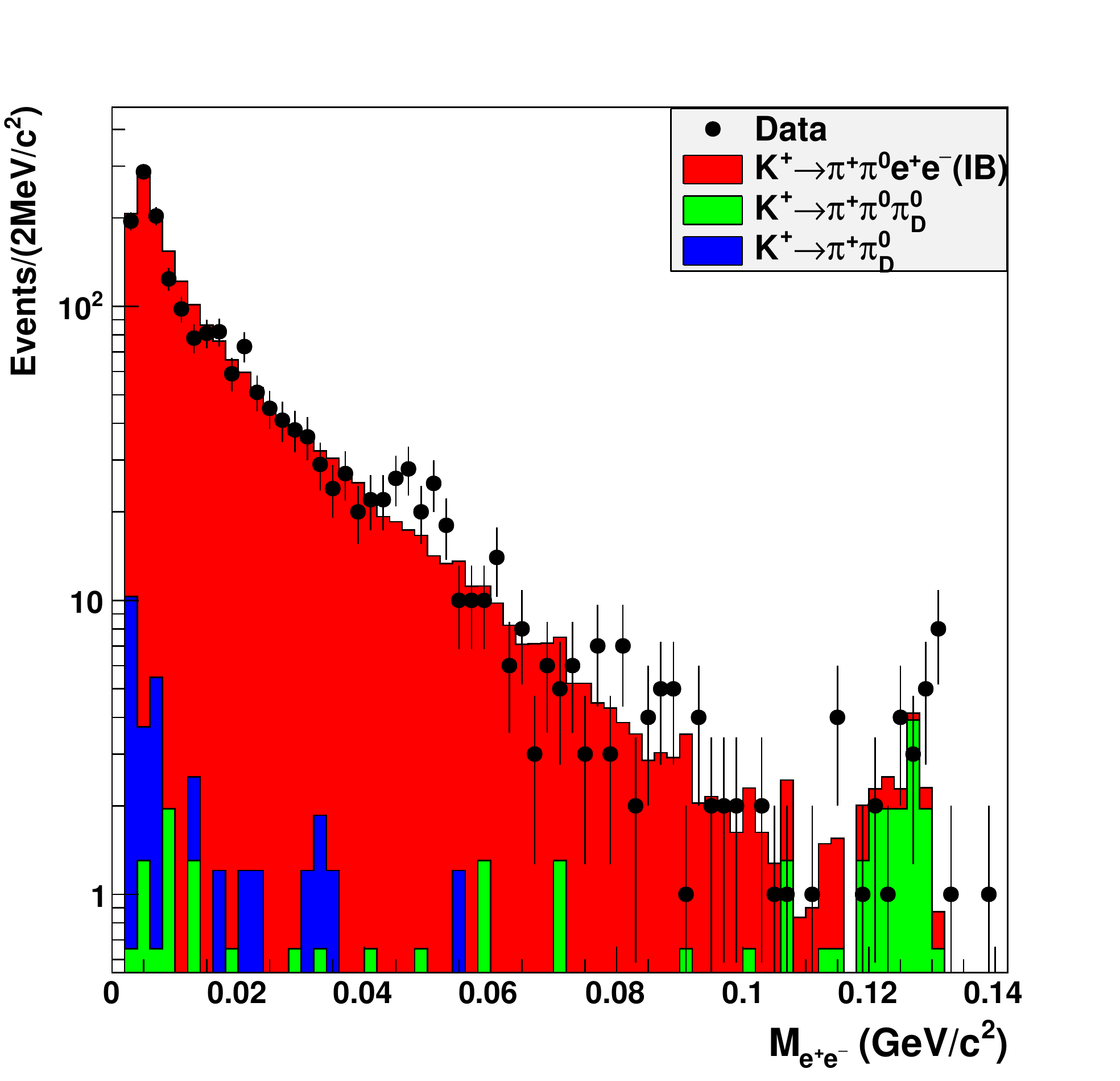}
        \caption{Reconstructed $e^+e^-$ invariant mass distributions of the data and simulated background samples.}
        \label{fig:ppeemee}
    \end{minipage}
\end{figure}
The MC simulation for the different \KPPEE contributions IB, DE, and the electric
interference, have been generated separately according to the theoretical description given in \cite{Cappiello:2011qc} 
neglecting the magnetic interference in the present preliminary result. 
Due to limited statistic of the data sample the extraction of the DE and electric interference is
not possible in this analysis.
The signal acceptance has been obtained from a weighted average of the single components acceptances, using as weights the relative fractions computed in \cite{Cappiello:2011qc} on the basis of the measurement of magnetic and electric terms of $K^{\pm} \to \pi^{\pm}\pi^{0}\gamma$ in \cite{Batley:2010aa}:
\begin{equation}
A_S=\frac{A_{IB}+A_{DE}\cdot Frac_{DE}+A_{INT}\cdot Frac_{INT}}{1+Frac_{DE}+Frac_{INT}}
\end{equation}
To take into account the E,M measurement uncertainties \cite{Batley:2010aa}, the weights entering the total  signal acceptance were varied accordingly  resulting in a $\sim 1\%$ relative change  quoted as the systematic uncertainty due to  the acceptance modeling.
As radiative corrections to the \KPPEE mode are not computed in \cite{Cappiello:2011qc}, 
the NA48/2 signal MC simulation included the following effects: the classical Coulomb attraction/repulsion between charged particles
and the real photon(s) emission as implemented in the PHOTOS package
The preliminary result for the total branching ratio is obtained:
\begin{equation}
{\cal B}(K^{\pm} \to \pi^{\pm}\pi^{0}e^+e^-)=(4.06\pm0.10_{stat.} \pm 0.06_{syst.} \pm 0.13_{ext.}) \cdot 10^{-6}
\end{equation} 
where systematic errors include uncertainties on acceptance, particle identification, trigger efficiencies and radiative corrections.
The external error originating from the normalisation mode branching ratio uncertainty is the dominant error in the present measurement
obtained with an overall precision of about 3\%. 
The comparison with theoretical expectations is presented in Figure \ref{fig:BRppee}. 
The small dashed blue line represents the theoretical prediction with no isospin breaking correction published in \cite{Cappiello:2011qc}. The
big dashed blue line shows the expected isospin breaking corrected branching ratio (private communication from the authors of \cite{Cappiello:2011qc}).
The experimental value of the ${\cal B}(K^{\pm} \to \pi^{\pm}\pi^{0}e^+e^-)$ is in a very good agreement with the theoretical predictions (within one standard deviation).
The NA48/2 data sample analyzed has no sensitivity to the DE and INT  contributions to the $M_{ee}$ spectrum within the current statistics (see Figure \ref{fig:ppeemee}).  It will be difficult to perform a full Dalitz plot analysis without a proper description of the radiative effects, particularly relevant  in a final state with two electron/positron.

\section{Search for the dark photon in $\pi^0$ decays}

The large sample of $\pi^0$ mesons produced and decaying in vacuum collected by NA48/2 allows for a high
sensitivity search for the dark photon ($A'$), a new gauge boson introduced in hidden sector new physics
models with an extra $U(1)$ gauge symmetry. In a rather general set of models, the interaction of the dark
photon (DP) with the ordinary matter is through kinetic mixing with the Standard Model hypercharge $U(1)$~\cite{ho86}. In these models, the
new coupling constant $\varepsilon$ is proportional to the electric charge and the dark photon couples in exactly the same way to quarks
and leptons. These scenarios could provide an explanation to the observed rise in the cosmic-ray positron fraction with energy, and could
offer an explanation to the muon gyromagnetic ratio ($g-2$) anomaly~\cite{po09}.
The simplest DP model is characterised by two free parameters, the DP mass $m_{A'}$ and the mixing parameter with the standard model
$\varepsilon$. Its possible production in the $\pi^0$ decay and subsequent decay proceed via the following chain:
$K^\pm\to\pi^\pm\pi^0$, ~~~ $\pi^0\to\gamma A'$, ~~~ $A'\to e^+e^-$,
producing a final state with three charged particles and a photon. The expected branching fraction of the $\pi^0$ decay is~\cite{batell09}:
\begin{equation}
{\cal B}(\pi^0\to\gamma A') = 2\varepsilon^2 \left(1-\frac{m_{A'}^2}{m_{\pi^0}^2}\right)^3 {\cal B}(\pi^0\to\gamma\gamma),
\label{eq:br}
\end{equation}
with a strong kinematic suppression of the decay rate for DP masses approaching $m_{\pi^0}$. In the mass range $2m_e\ll m_{A'}<m_{\pi^0}$
accessible in this analysis, assuming that the DP can only decay into SM fermions, ${\cal B}(A'\to e^+e^-)\approx 1$ while the allowed
loop-induced decays
($A'\to 3\gamma$, $A'\to\nu\bar\nu$) are highly suppressed. 
The maximum DP mean path~\cite{batell09} in the NA48/2 experiment corresponds to an energy of approximately $E_{\rm max}=50~{\rm GeV}$:
\begin{displaymath}
L_{\rm max} \approx (E_{\rm max}/m_{A'}) c\tau \approx 0.4~{\rm mm} \times \left(\frac{10^{-6}}{\varepsilon^2}\right) \times \left(\frac{100~{\rm MeV}}{m_{A'}}\right)^2,
\end{displaymath}
In the accessible parameter range ($m_{A'}>10$~MeV/$c^2$ and $\varepsilon^2>5\times 10^{-7}$) $L_{\rm max}$ does not exceed 10~cm and the DP
can be assumed to decay at the production point. In this prompt decay scenario the NA48/2 3-track vertex reconstruction does not introduce
significant acceptance losses as the typical resolution on the vertex longitudinal coordinate is $\approx 1$~m. The DP signature is
identical to that of the Dalitz decay $\pi^0_D\to e^+e^-\gamma$, which therefore represents an irreducible background and limits the
sensitivity. The largest $\pi^0_D$ sample, and therefore the largest sensitivity, is obtained form the study of the $K^\pm\to\pi^\pm\pi^0_D$ decays.

\subsection{Event selection and background simulation}
\label{sec:selection}

\begin{figure}[ht]
    \centering
    \begin{minipage}[t]{0.45\textwidth}
        \centering
        \includegraphics[width=60mm,scale=0.5]{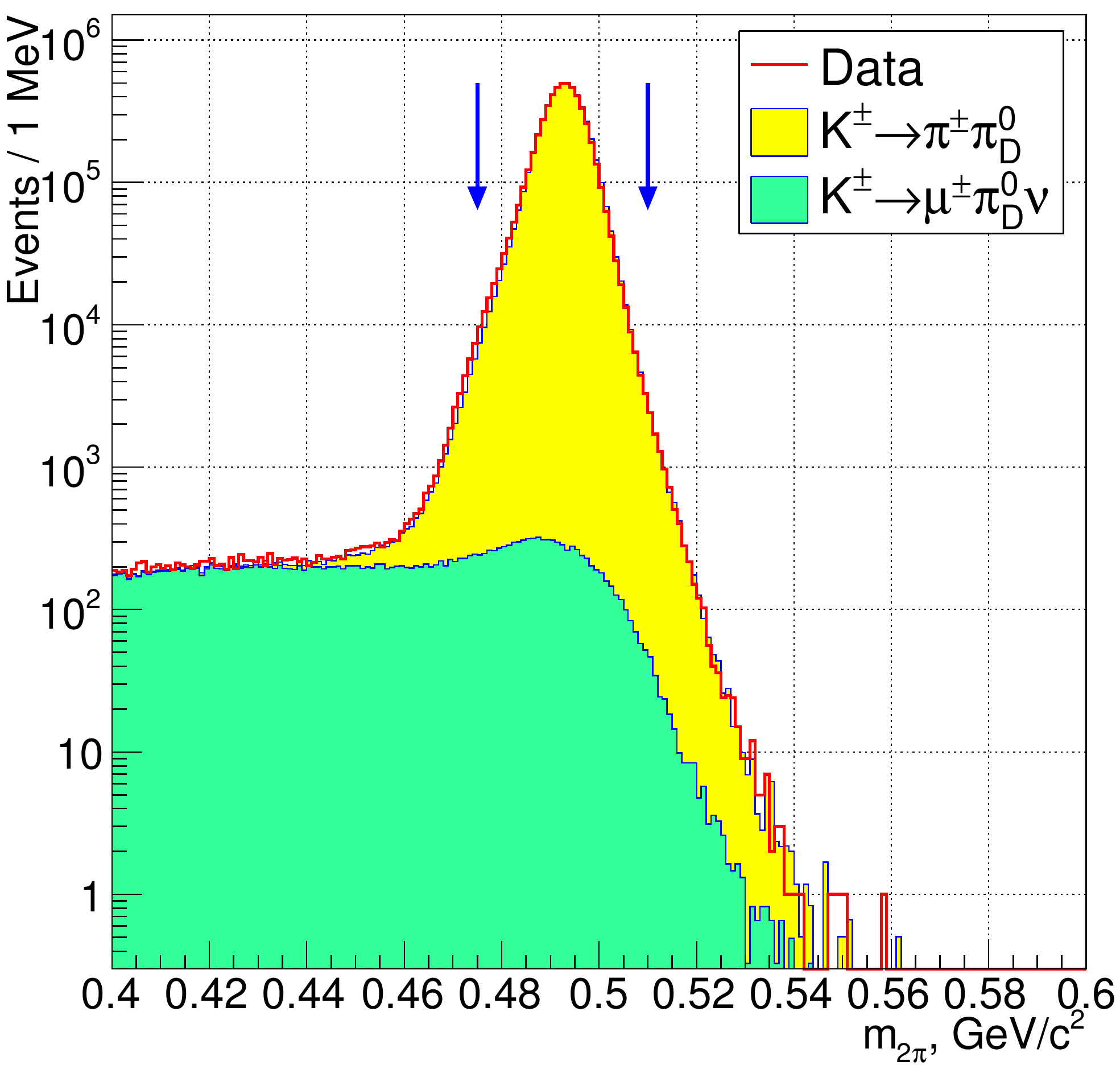}%
        \caption{Reconstructed $\pi^\pm\pi^0_D$ invariant mass ($m_{2\pi}$) distributions of the data and simulated background samples. The selection condition is illustrated with arrows.}
        \label{fig:mass}
    \end{minipage}%
    \hspace{1cm}
    \begin{minipage}[t]{0.45\textwidth}
        \centering
        \includegraphics[width=60mm,scale=0.5]{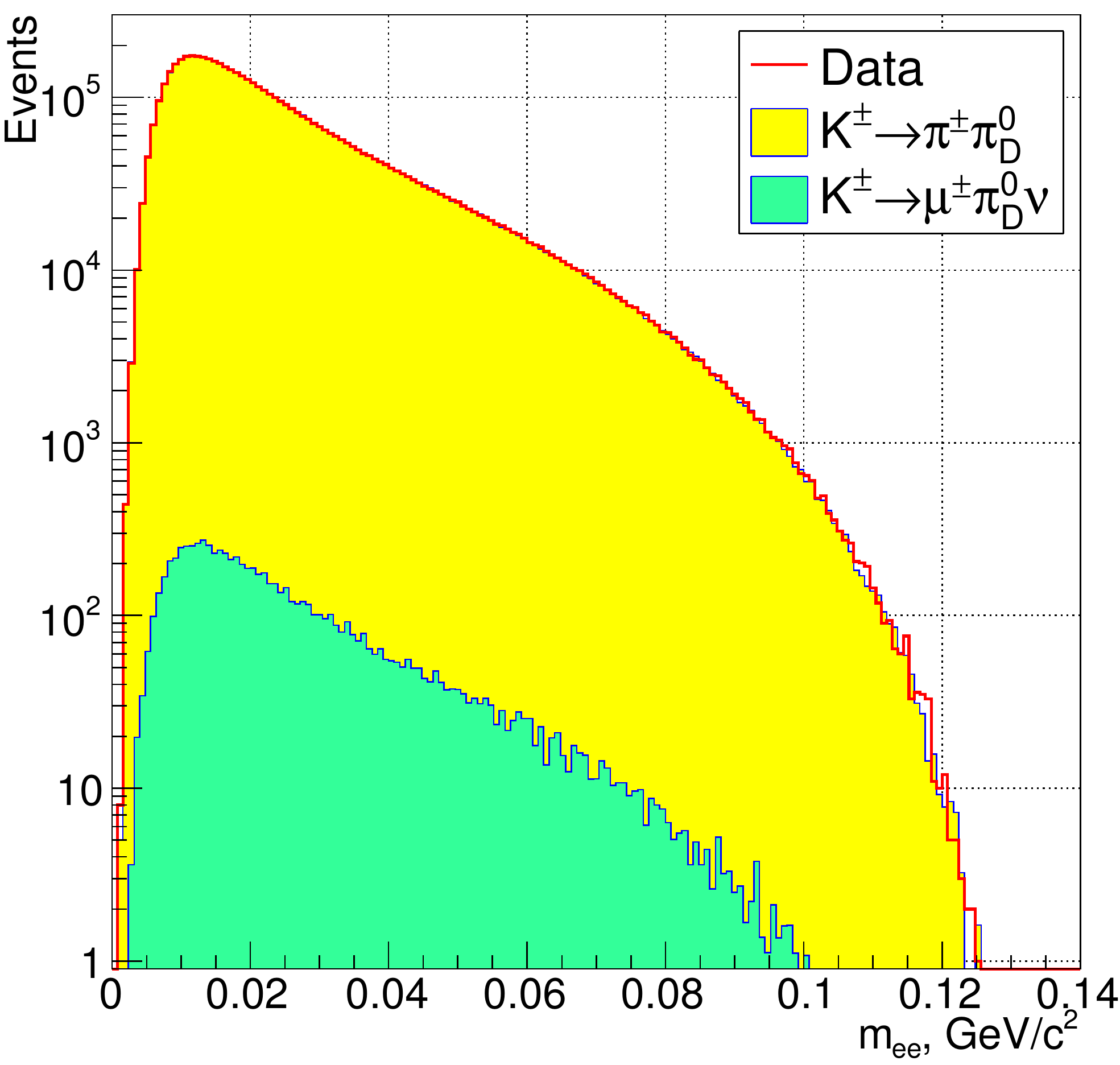}
        \caption{Reconstructed $e^+e^-$ invariant mass distributions of the data and simulated $K_{2\pi D}$ and $K_{\mu 3D}$ samples.}
        \label{fig:mee}
    \end{minipage}
\end{figure}

The full NA48/2 data sample is used for the analysis. The $K_{2\pi D}$ event selection requires a three-track vertex reconstructed in the
fiducial decay region formed of a pion ($\pi^\pm$) candidate track and two opposite-sign electron ($e^\pm$) candidate tracks. Charged
particle identification is based on the ratio of energy deposition in the LKr calorimeter to the momentum measured by the spectrometer,
which should be smaller (greater) than 0.85 for pion (electron) candidates. Furthermore, a single isolated LKr energy deposition cluster
is required as the photon candidate. 
The reconstructed invariant mass of the $\pi^\pm\pi^0$ system (Fig.~\ref{fig:mass}) is required to the consistent with the $K^\pm$ mass.
A sample of $4.687\times 10^6$ fully reconstructed $\pi^0_D$ decay candidates in the $e^+e^-$ invariant mass range $m_{ee}>10~{\rm MeV}/c^2$
with a negligible background is selected. The candidates mainly originate from $K_{2\pi D}$ decays, with 0.15\% coming from the semileptonic
$K^\pm\to\pi^0_D\mu^\pm\nu$ decays (denoted $K_{\mu 3 D}$ below). Correcting the observed number of candidates for acceptance and trigger
efficiency, the total number of $K^\pm$ decays in the 98~m long fiducial decay region for the analyzed data sample is found to be
$N_K=(1.55\pm0.05)\times 10^{11}$, where the quoted error is dominated by the external uncertainty on the $\pi^0_D$ decay branching
fraction ${\cal B}(\pi^0_D)$. The reconstructed $e^+e^-$ invariant mass ($m_{ee}$) spectrum of the $K_{2\pi D}$ candidates is displayed
in Fig.~\ref{fig:mee}. A dark photon produced in the $\pi^0_D$ decay and decaying promptly to $e^+e^-$ would appear as narrow peak in the
spectrum. Monte Carlo (MC) simulations of the $K_{2\pi D}$ and $K_{\mu 3D}$ processes are performed to subtract the irreducible $\pi^0_D$
background. The $\pi^0_D$ decay is simulated using the lowest-order differential decay rate in~\cite{mi72}.
Radiative corrections to the differential rate are implemented following the approach of Mikaelian and Smith~\cite{mi72} recently revised
to improve the numerical precision~\cite{Husek}. 
The method introduces only weights $\delta(x,y)$ and does not account for the emission of inner bremsstrahlung photons.
\subsection{Dark photon search technique}
\label{sec:dp}
A search for the DP is performed assuming different mass hypotheses with a variable mass step. The mass step of the scan and the width of
the signal mass window around the 
assumed DP mass are determined by the resolution on the $e^+e^-$ invariant mass. The mass step of the DP scan is set to be $\sigma_{m}/2$,
while the signal 
region mass window for each DP mass hypothesis is defined as $\pm1.5\sigma_m$ around the assumed mass (both the scan step and the mass
window half-width are rounded 
to the nearest multiple of 0.02 MeV/$c^2$). The mass window width has been optimised with MC simulations to obtain the highest sensitivity
to the DP signal, determined by a 
trade-off between $\pi^0_D$ background fluctuation and signal acceptance.

\begin{figure}[ht]
    \centering
    \begin{minipage}[t]{0.45\textwidth}
    \centering
         \includegraphics[width=70mm,scale=0.8]{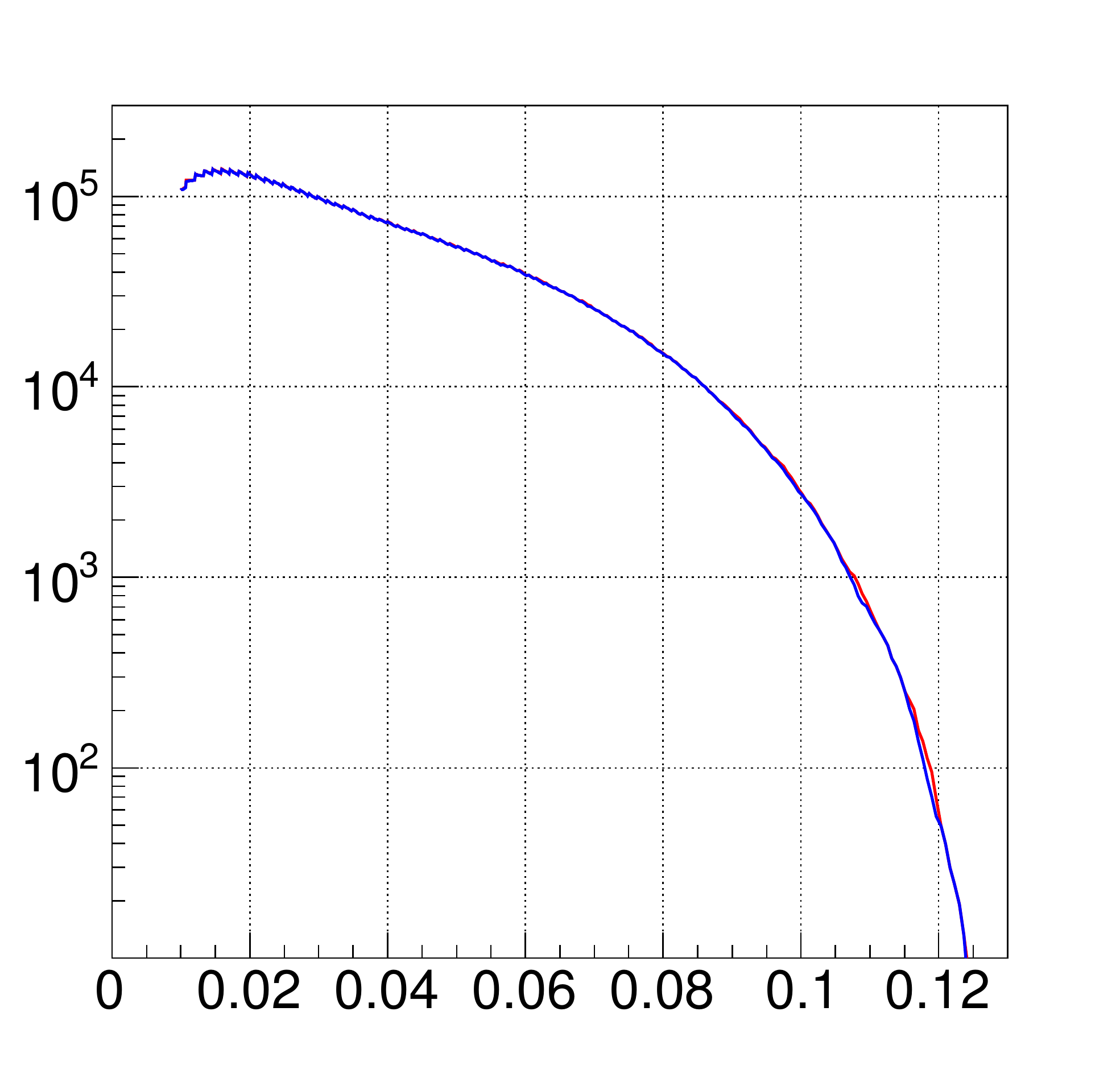}
         \caption{Numbers of observed (red) and expected (blue) DP candidates in the DP signal mass window as for each assumed DP mass hypothesis.}
	\label{fig:observed}
    \end{minipage}%
    \hspace{1cm}
    \begin{minipage}[t]{0.45\textwidth}
       \centering
	\includegraphics[width=65mm,scale=0.8]{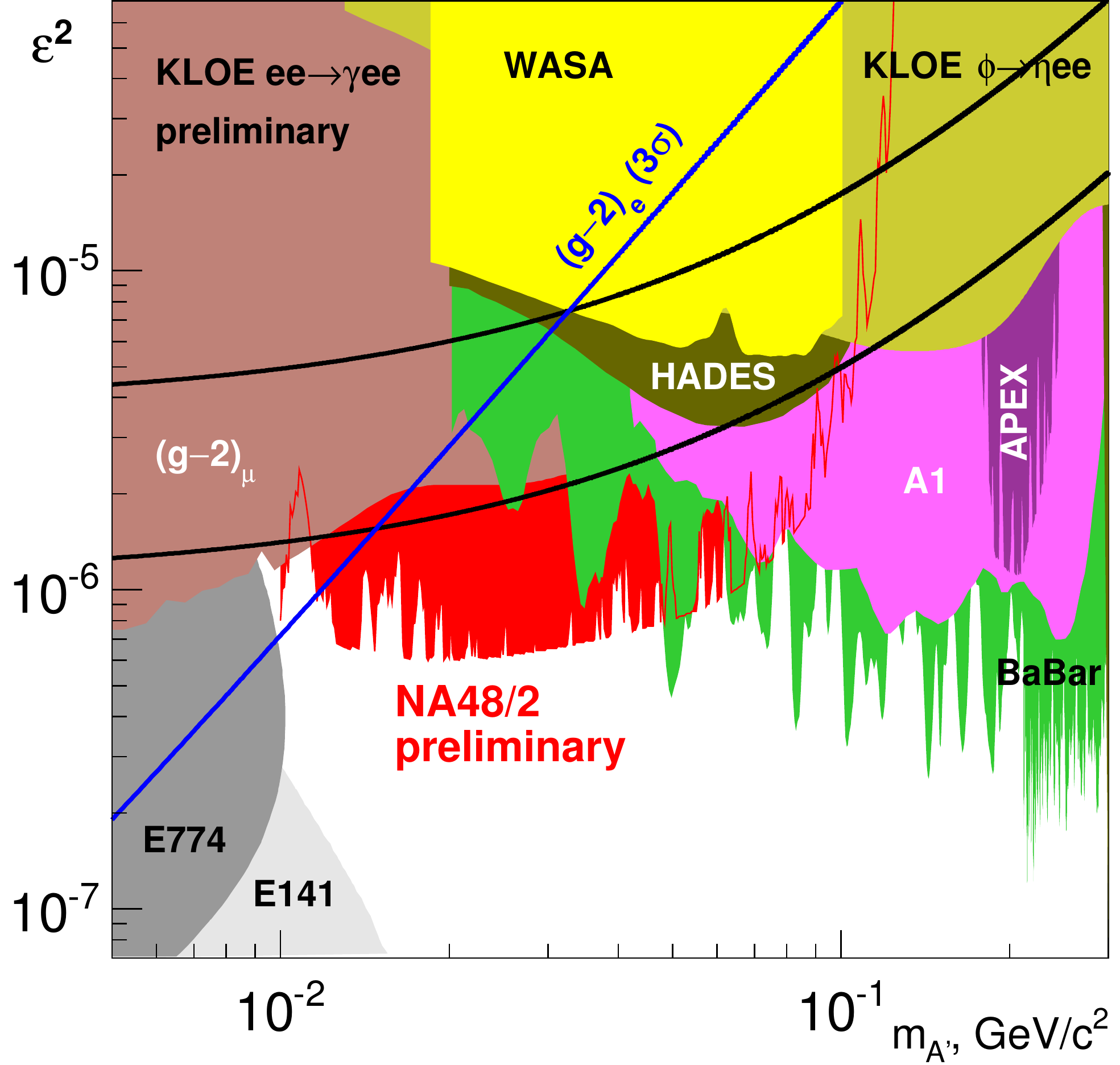}
	\caption{The NA48/2 preliminary upper limits at 90\% CL on the mixing parameter $\varepsilon^2$ versus the DP mass $m_{A'}$, compared to the other published exclusion 	limits.}
	\label{fig:world}
    \end{minipage}
\end{figure}

In total, 398 DP mass hypotheses are tested in the range $10~{\rm MeV}/c^2 \le m_{ee} < 125~{\rm MeV}/c^2$. The lower limit of the
considered mass range is determined by  the limited precision of MC simulation of background at low mass, while at the upper limit of the
mass range the signal acceptance drops to zero. The numbers of observed data 
events in the signal region ($N_{\rm obs}$) and the numbers of $\pi^0_D$ background events expected from MC simulation corrected by the
measured trigger efficiencies 
($N_{\rm exp}$) in the DP signal window for each considered mass hypothesis are presented in Fig.~\ref{fig:observed}. They decrease with
the DP mass due to the steeply falling 
$\pi^0_D$ differential decay rate and decreasing acceptance, even though the mass window width increases, being approximately proportional
to the mass.

Confidence intervals at 90\% CL for the number of $A'\to e^+e^-$ decay candidates ($N_{\rm DP}$) in each mass hypothesis ($N_{\rm DP}$) are
available from $N_{\rm obs}$, 
$N_{\rm exp}$ and $\delta N_{\rm exp}$ using the Rolke--L\'opez method~\cite{rolke} assuming Poissonian (Gaussian) errors on the numbers of
observed (expected) events. For 
the preliminary results, it is assumed conservatively that $N_{\rm obs}=N_{\rm exp}$ in cases when $N_{\rm obs}<N_{\rm exp}$, as the employed
implementation of the method 
(from the ROOT package) has been found to underestimate the upper limits in that case.
Upper limits at 90\% CL on ${\cal B}(\pi^0\to\gamma A')$ in each DP mass hypothesis in the assumption ${\cal B}(A'\to e^+e^-)=1$ are
computed using the relation
\begin{displaymath}
{\cal B}(\pi^0\to\gamma A') = \frac{N_{\rm DP}}{N_K}
\Big[{\cal B}(K_{2\pi}) A(K_{2\pi}) + {\cal B}(K_{\mu 3}) A(K_{\mu 3}) \Big]^{-1}.
\end{displaymath}
The acceptances $A(K_{2\pi})$ and $A(K_{\mu 3})$ of the employed $K_{2\pi D}$ event selection for the $K_{2\pi}$ and $K_{\mu 3}$ decays,
respectively, followed by the prompt 
$\pi^0\to\gamma A'$, $A'\to e^+e^-$ decay chain, are evaluated for each considered DP mass with MC simulation. Event distributions in the
angle between $e^+$ momentum in 
the $e^+e^-$ rest frame and the $e^+e^-$ momentum in the $\pi^0$ rest frame are identical for the decay chain involving the DP
($\pi^0\to\gamma A'$, $A'\to e^+e^-$) and the $
\pi^0_D$ decay, up to a negligible effect of the radiative corrections that should not be applied in the former case. Therefore DP
acceptances are evaluated using the MC samples produced for background description.

The largest uncertainty on the computed ${\cal B}(\pi^0\to\gamma A')$ is the external one due to ${\cal B}(\pi^0)$ entering via $N_K$.
It amounts to 3\% in relative terms and is
neglected. The obtained upper limits on ${\cal B}(\pi^0\to\gamma A')$ are ${\cal O}(10^{-6})$ and do not exhibit a strong dependence on
the assumed DP mass, as the negative
trends in background fluctuation (Fig.~\ref{fig:observed}) and acceptance largely cancel out.
Upper limits at 90\% CL on the mixing parameter $\varepsilon^2$ in each considered DP mass hypothesis are calculated from those on
${\cal B}(\pi^0\to\gamma A')$ using Eq.~(\ref{eq:br}). 
The resulting preliminary DP exclusion limits, along with constraints from other experiments~\cite{le14}, the band of phase space where the
discrepancy between the measured 
and calculated muon $g-2$ values falls into the $\pm2\sigma$ range~\cite{po09,da14} due to the DP contribution, and the region excluded by
the electron $g-2$ measurement, 
are presented in Fig.~\ref{fig:world}. The obtained upper limits on $\varepsilon^2$ represent an improvement over the existing data in the
DP mass range 10--60~MeV/$c^2$.
Under the assumption that the DP couples to SM through kinetic mixing and decays predominantly to SM particles, the NA48/2 preliminary
result excludes the DP as an explanation for the muon (g-2) anomaly in the range 10-100 MeV/$c^2$.

\end{document}